\documentclass[%
 reprint,
superscriptaddress,
preprintnumbers,
 amsmath,amssymb,
 aps,nofootinbib
]{revtex4-1}

\usepackage{graphicx}
\usepackage{dcolumn}
\usepackage{bm}
\usepackage{hyperref}
\usepackage{color}
\usepackage{xcolor}
\usepackage{soul}
\usepackage{feynmp-auto}
\usepackage{float}

\newcommand{\f}{\psi}   
\newcommand{\be}{\begin{equation}}
\newcommand{\ee}{\end{equation}}
\newcommand{\bear}{\begin{eqnarray}}
\newcommand{\eear}{\end{eqnarray}}

\begin{document}

\preprint{ FERMILAB-PUB-23-661-T}
\title{Mutually elusive: vectorlike antileptons and leptoquarks}

\author{Innes Bigaran}
 \affiliation{Department of Physics \& Astronomy, Northwestern University, 2145 Sheridan Road, Evanston, IL 60208, USA}
\affiliation{Particle Theory Department, Fermilab, P.O. Box 500, Batavia, IL 60510, USA}

\author{Bogdan A. Dobrescu}
\affiliation{Particle Theory Department, Fermilab, P.O. Box 500, Batavia, IL 60510, USA}

\author{Alessandro Russo}
\affiliation{ETH Zürich
CH-8093 Zürich, Schafmattstrasse 16, Zurich, Switzerland  \\  { $\,$} }

\vspace*{0.3cm}

\date{December 19, 2023; revised March 21, 2024}   

\begin{abstract}
We study the properties of vectorlike fermions that have the same gauge charges as the Standard Model lepton doublets, but opposite lepton number. These antileptons undergo decays mediated by heavier scalar leptoquarks, while the symmetries of this renormalizable model protect the vectorlike fermions and the leptoquarks from standard decays probed so far at colliders. We derive upper limits on the new Yukawa couplings imposed by flavor-changing processes, including $B \to K \nu \bar \nu$ and $B_s-\overline B_s$ mixing, and show that they are 
compatible with prompt antilepton decays at the LHC for wide parameter ranges. If the new particles couple predominantly to second-generation quarks, then their collider probes involve multiple jets and two taus or neutrinos, and are hampered by large backgrounds. If couplings to third-generation quarks are large, then the collider signals involve top quarks, and can be probed more efficiently at the LHC. Even in that case, both the vectorlike fermion doublet and the leptoquarks remain more elusive than in models with standard decays. \\
\end{abstract}


\maketitle


\section{Introduction}

Chirality is an essential characteristic of all Standard Model~(SM) fermions. 
The left- and right-handed components  of the SM fermions transform differently under the electroweak gauge group, and thus the quarks and leptons acquire masses by coupling to the Higgs field.
Vectorlike fermions are distinct from chiral fermions in that the transformation properties under the SM gauge group are identical for both chiralities.
A Dirac mass for a vectorlike fermion does not violate any SM symmetry. Therefore, the size of their masses is not tied to the Higgs doublet.
If elementary vectorlike fermions exist, they represent a new form of matter, and dedicated searches for these are a major goal for the LHC experiments.

A vectorlike fermion that is a color singlet and carries lepton number $L = +1$ is referred to as a vectorlike lepton. It is typically assumed that  the vectorlike leptons have the same charges under the $ SU(3)_c\otimes SU(2)_W\otimes U(1)_Y$ gauge group as a SM lepton, 
namely $(1,2,-1/2)$ for a weak doublet, or $(1,1,-1)$ for a weak singlet. In that case, a Yukawa coupling to the SM Higgs doublet and a SM lepton field is allowed by the gauge symmetry, and induces a mass mixing between the vectorlike lepton and the SM lepton \cite{Falkowski:2013jya,Kumar:2015tna,Bhattiprolu:2019vdu}). As a result the vectorlike lepton decays into a SM lepton and a $W$, $Z$, or Higgs boson. Searches for electroweak production of vectorlike leptons that decay  
under these hypotheses have been performed by the CMS collaboration \cite{CMS:2019hsm,CMS:2022nty},  and a particularly stringent lower mass limit, of 1.05 TeV, has been set on the vectorlike lepton doublet.

In this article, we study the properties of a vectorlike fermion $\f$ that transforms as $(1,2,-1/2)$ under the SM gauge group, but which carries lepton number $L = -1$. 
As this is the opposite of the lepton number carried by the SM leptons, we generically refer to $\f$ as a Vectorlike AntiLepton (VAL). 
 We assume that  $\f$ has a Yukawa interaction to a heavier leptoquark scalar ($\xi$) and a SM quark, and that all renormalizable interactions of $\f$ preserve lepton number.  Consequently, 
 the coupling of $\f$ to any SM lepton and the Higgs doublet is forbidden. Thus, the current searches for vectorlike leptons do not apply to $\f$, as there are no standard decays ({\it i.e.}, into a SM lepton and a SM boson). Instead, electroweak production of a $\f$ pair is followed, in our model,  by 3-body decays mediated by the leptoquark. 
 
The reverse is also true: instead of decaying into a SM lepton  and a quark (for a review, see \cite{Dorsner:2016wpm}),  the leptoquark predominantly decays into a VAL and a SM quark, leading to a cascade decay.
As a result, the VAL and the scalar leptoquark mutually elude present searches for vectorlike leptons or leptoquarks at the LHC, and motivate new dedicated searches.

Certain elements of the renormalizable model proposed here have been encountered in other contexts. 
Three-body decays of vectorlike quarks via a heavy scalar leptoquark have been analyzed in \cite{Dobrescu:2016pda}.
Three-body decays of a vectorlike lepton through a leptoquark of spin-1 have been studied in \cite{DiLuzio:2017vat,Bordone:2017bld, DiLuzio:2018zxy,Greljo:2018tuh}, leading to $\tau b \bar b$ final states searched for at CMS~\cite{CMS:2022cpe}; 
we will show that this is different than the final states obtained in the case of the VAL. Nonstandard decays of a vectorlike lepton, into a SM lepton and a light pseudoscalar, are discussed in \cite{Bernreuther:2023uxh} (see also \cite{Cao:2023smj}); in that model, though, the standard decays remain important.

On a more general note, vectorlike fermions appear in various frameworks beyond the SM, including bound states in models of quark and lepton compositeness \cite{Dobrescu:2021fny}, 
or higher multiplets of grand unified groups \cite{Hewett:1988xc}, so  if they exist, vectorlike fermions would likely provide a window towards deeper principles.
Thus, it is important to analyze a wide range of possible properties of vectorlike fermions, and to search for a variety of signals in experiments. The VALs studied here are as motivated as the vectorlike leptons that have mass mixing with the SM leptons, and thus are worth exploring in more detail.

In Section \ref{sec:Model} we present the interactions of the VAL and of the leptoquark,  analyze constraints from flavor processes, and compute the decay length of $\psi$. 
In Section \ref{sec:pheno} we study the LHC phenomenology of these particles, focusing on signals involving quarks of either second or third generation.
We summarize our results in Section \ref{sec:conc}, where we also list the most important LHC searches that can test this model.


\section{Vectorlike antilepton doublet} 
\label{sec:Model}
\setcounter{equation}{0}

Let us consider an extension of the SM that includes a weak-doublet vectorlike fermion $\f =( \f^{0}, \f^{-})^{\top}$ 
that carries lepton number $L= -1$ and transforms under the $SU(3)_c\otimes SU(2)_W \otimes U(1)_Y$ gauge group 
as $(1, 2, -1/2)$. Since the SM lepton doublets $\ell^j_{_L}$ ($j=1,2,3$ is the generation index) have the same gauge charges as $\f$,
but opposite lepton number (by convention the electron has lepton number $+1$), it is appropriate to refer to $\f$ as an antilepton. 
A Dirac mass term for the $\f$ doublet ($- m_\f \, \overline \f_{\! _L} \, \f_{\! _R} + {\rm H.c.}$)
is allowed both by the SM gauge symmetry (since the left- and right-handed components have the same charges) and by lepton number conservation. 

We assume that the interactions of $\f$ preserve lepton number.
Thus, any mixing between $\f$ and the SM leptons can be safely neglected, because it is forbidden by lepton number conservation. 
Note that the LEP searches for new charged particles imply $m_\f \gtrsim 100$ GeV \cite{L3:2001xsz, ALEPH:2013htx}.
Even if the SM neutrinos acquire Majorana masses (which is not necessary, as the neutrinos may have Dirac masses), 
the lepton number breaking effects are suppressed by powers of $m_\nu/m_\f \lesssim 10^{-12}$. 

Although both the left- and right-handed components of $\f$ have $L= -1$, one could use instead their charge-conjugate fields ($\f_{\! _L}^c$ and $\f_{\! _R}^c$) which have $L= +1$. Thus,  $\f^c$ may be viewed as a vectorlike lepton of wrong-sign hypercharge (compared to the SM lepton doublets), so the mixing between $\f^c$ and the SM leptons is still forbidden. 

In the absence of additional fields or of higher-dimensional operators, the lightest particle of the $\f$ doublet remains stable.
We introduce, however, a leptoquark scalar $\xi$ that transforms as $(3,2,+1/6)$ under 
$SU(3)_c\otimes SU(2)_W \otimes U(1)_Y$ that carries lepton number $L= -1$. Gauge and Lorentz invariances allow the following 
Yukawa interaction of $\xi$ with $\f$ and with the SM up-type quark singlets (labeled $u_{j_R}$ in the gauge eigenstate basis):
\be
\lambda_{j _\f} \,  \xi^\dagger \, \overline{u}^{\,  c}_{j_R} \, \f_{\! _R} + {\rm H.c.}
\label{eq:Fconserving}
\ee
Here $\lambda_{j _\f} $ are three dimensionless couplings. 
If the leptoquark mass satisfies $M_\xi > M_\f$, then the above Yukawa interactions still do not allow the decay of the lightest particle of the $\f$ doublet. This is due to a global $U(1)_\f$ symmetry under which both $\f$ and $\xi$ carry charge +1, while the SM fields are singlets. 

Renormalizable gauge-invariant terms that explicitly break  $U(1)_\f$ can be written as
\be
\lambda_{j \tau} \,  \xi \; \overline{d}_{j_R} \, \ell^3_{\! _L} + {\rm H.c.}
\label{eq:Fbreaking}
\ee
We assume that the above $U(1)_\f$ breaking is suppressed, so the Yukawa couplings satisfy $\lambda_{j \tau} \ll \lambda_{j _\f}$.
The Yukawa interaction (\ref{eq:Fbreaking}) involves only the third-generation lepton doublet because we impose that $\tau$-lepton number ($L_\tau $) is conserved in this model, assigning $L_\tau = -1$ to both $\f$ and $\xi$.
Focusing on the interaction with $\ell^3_{\! _L} = (\nu_{\tau_L}, \tau_{_L})$ is motivated in part by the larger $\tau$ Yukawa coupling to the Higgs doublet in the SM, which suggests that new physics might also have larger couplings to $\tau$ than to $\mu$ or $e$.
A consequence of this coupling is that we will focus on collider signatures containing taus, which have larger backgrounds than those with final states containing muons or electrons. 

Assigning baryon number $B = 1/3$ to $\xi$, and $B = 0$ to $\f$, ensures that $B$ is preserved by all the interactions introduced here.
We point out that scalars that transform as $(3, 2, +1/6)$ under the SM gauge group have been studied in various contexts, such as 
squarks (labeled $\tilde q$) in supersymmetric extensions of the SM, or leptoquark doublets (sometimes labeled $\tilde R_2$ \cite{Dorsner:2016wpm}
or $\Phi_{\tilde 2}$ \cite{Crivellin:2021ejk}). We use here the notation $\xi$ to emphasize that the scalar participates in the peculiar interaction  (\ref{eq:Fconserving}). 

Besides the gauge interactions of $\psi$ and $\xi$, and the Yukawa interactions discussed above, the model includes a single renormalizable interaction, namely  $\lambda_{ _H} \, \xi^\dagger \xi H^\dagger H$.
This Higgs portal coupling for the leptoquark changes both the Higgs production via gluon-fusion and the Higgs decay widths to $\gamma\gamma$ and $\gamma Z$ (see \cite{Dorsner:2016wpm} for a review). 
However, the ensuing constraints on the $\lambda_{ _H}$ coupling remain relatively mild for now \cite{Crivellin:2020ukd}.


\subsection{Constraints from flavor processes}

In the mass eigenstate basis for the up-type quarks, the Yukawa interactions (\ref{eq:Fconserving}) take the form
\be
\xi^\dagger  \left(   \lambda_{t} \, \overline{t}^{\,  c}_{ _R}  +  \lambda_{c} \, \overline{c}^{\,  c}_{ _R}  +  \lambda_{u} \, \overline{u}^{\,  c}_{ _R}  \right)  \f_{\! _R} ~~.
\label{eq:FconservingMass}
\ee
Here and in what follows it is implicit that all the interaction terms are accompanied by their Hermitian conjugates in the Lagrangian.
Note that there is a linear relation between the Yukawa couplings introduced in (\ref{eq:Fconserving}) and  (\ref{eq:FconservingMass}) through the $3\times 3$ matrix $U_R$ that relates the physical and gauge eigenstate bases for the right-handed up-type quarks. As $U_R$ is not determined by any existing measurements, we focus on the physical couplings $\lambda_{t}, \lambda_{c}, \lambda_{u}$ instead of $\lambda_{j _\f}$. 

\begin{figure}[t]
\centering 
\vspace*{.5mm}
\includegraphics[scale=1.06]{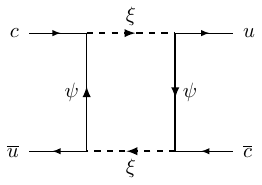}
\vspace*{-0.3cm}
 \caption{Box diagram contributing to $D^0 - \overline{D^0}$ meson mixing,  with the $\f$ VAL and $\xi$ scalar running in the loop.}
\label{fig:DDbar}
\end{figure}

The terms involving the up and charm quarks give rise to FCNC processes involving $D$ mesons. In particular, there are contributions
to $D^0 - \overline{D^0}$ meson mixing from box diagrams with $\f$ and $\xi$ are running in the loop, such as the one shown in Figure~\ref{fig:DDbar}.
The $\Delta C = 2$ operator generated by the box diagrams is given by 
\be
f(m_{ _\f} /M_\xi)  \, \frac{\lambda_{c}^2 \, \lambda_{u}^{*2}}{16 \pi^2\,  M_\xi^2} \, 
(\overline u_{ _R}^\alpha c_{ _L}^\beta) (\overline u_{ _R}^\beta c_{ _L}^\alpha )  ~~,
\ee
where $\alpha, \beta$ are color indices, and  $f(m_{ _\f} /M_\xi) $ is a function with values roughly of order one. 
A discussion of the above operator, labeled $Q_3$, can be found in \cite{Silvestrini:2019sey, Gabbiani:1996hi}.  
The comparison of the measured $D^0 - \overline {D^0}$ mixing with the SM prediction (which suffers from large QCD uncertainties related to the charm quark) imposes that the coefficient 
of the $(\overline c_{ _L} u_{ _R})^2$ operator has a real part smaller than $6 \times 10^{-8}$ TeV$^{-2}$ and an imaginary part smaller by an extra factor of about 5 \cite{Isidori:2013ez}. Based on that constraint, we estimate
\be
\left\{ {\rm Re}( \lambda_{c} \, \lambda_{u}^*) \; , \, {\rm Im}(\lambda_{c} \, \lambda_{u}^*) \rule{0mm}{3.6mm}  \right\} 
\lesssim \left\{ O(3) \, , \, O(1) \rule{0mm}{3.6mm}  \right\}   \frac{ 10^{-3} \, M_\xi }{1 \; \rm TeV}  ~.
\label{eq:lambdacu}
\ee
In what follows we assume $\left| \lambda_{u} \right| \ll \left| \lambda_{c}  \right|$, so we neglect any effects due to $ \lambda_{u}$. Thus, we do not need a more precise estimate of the above constraint.

Searches for rare top-quark decays can in principle constrain the product 
$\left| \lambda_{t} \, \lambda_{c} \right|$. For example, the $t \to c g$ and $t \to c \gamma$ decays would be induced by $\psi$ and $\xi$ running in a loop. 
This constraint is weak given the current sensitivity of LHC experiments, because the new $t$ branching fractions are suppressed by both the loop factor and the large $\xi$ mass.

\begin{figure}[t]
\centering 
\vspace*{.5mm}
\includegraphics[scale=1.06]{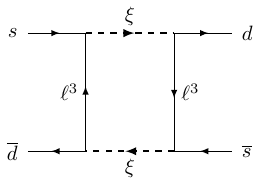}
\vspace*{-0.3cm}
 \caption{Box diagram contributing to $K^0 - \overline{K^0}$ meson mixing,  with the $\xi$ scalar leptoquark and the SM third-generation lepton doublet $\ell^3$ running in the loop. \\ }
\label{fig:KKbar}
\centering \vspace*{.5mm}
\includegraphics[scale=1.1]{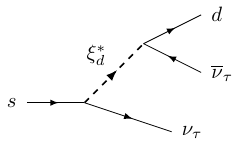}
\vspace*{-0.3cm}
\caption{Tree-level $s\to d \, \nu_\tau \, \overline \nu_\tau$ transition, which induces  $K \rightarrow \pi \, \nu \, \overline \nu$ decays, mediated by an off-shell leptoquark of electric charge $-1/3$. The width of this transition depends on the Yukawa couplings of the $\xi$ doublet to the $s_{ _R}$ and $d_{ _R}$ quarks.}
\label{fig:KpinunuDiagram}
\end{figure}

The Yukawa interactions of $\xi$ to the quarks of charge $-1/3$, shown in (\ref{eq:Fbreaking}), can be written in the mass eigenstate basis for the down-type quarks as
\be
\xi \left(   \lambda_{b} \, \overline{b}_{ _R}  +  \lambda_{s} \, \overline{s}_{ _R}  +  \lambda_{d} \, \overline{d}_{ _R}  \right)  \ell^3_{\! _L} ~~.
\label{eq:FbreakingMass}
\ee
Any of the three pairs of down-type quark flavors in these interactions is constrained by flavor-changing neutral processes. 
For example, there are contributions to $K^0 - \overline{K^0}$ meson mixing from box diagrams (one of them shown in Figure~\ref{fig:KKbar}) with $\xi = (\xi_u, \xi_d)^\top$  and the SM lepton doublet $\ell^3_{\! _L}$ running in the loop.
This process imposes constraints on the real and imaginary parts of $\lambda_{s} \, \lambda_{d}^*$ similar to (\ref{eq:lambdacu}), but stronger \cite{Mandal:2019gff}:  
\bear
&& \left\{ \left| {\rm Re}( \lambda_{s} \, \lambda_{d}^*)  \right| \; , \,  \left|  {\rm Im}(\lambda_{s} \, \lambda_{d}^*)  \right|  \rule{0mm}{3.6mm} \right\} 
\lesssim  \left\{ 3 \times \! 10^{-2} , \, 
10^{-4}  \rule{0mm}{3.6mm}  \right\}   \frac{ M_\xi }{1 \; \rm TeV}   .
\nonumber \\
&&
\label{eq:lambdasd}
\eear
Another set of flavor constraints arises from the tree-level $s\to d \, \nu_\tau \, \overline \nu_\tau$ transition mediated by the leptoquark of electric charge $-1/3$ ({\it i.e.}, $\xi_d$), as shown in Figure \ref{fig:KpinunuDiagram}. This contributes to the $K^+ \rightarrow \pi^+ \, \nu \, \overline \nu$ decay, and imposes  limits \cite{Mandal:2019gff}  on the real and imaginary parts of  
$\lambda_{s} \, \lambda_{d}^*$ that have a different $M_\xi$ dependence than the limits from $K^0 - \overline{K^0}$ mixing.
Conservatively the $K^+ \rightarrow \pi^+ \, \nu \, \overline \nu$ limits can be written as 
\be
\left| \lambda_{s} \, \lambda_{d}  \right|
\lesssim 4 \times 10^{-4}  \left( \frac{ M_\xi }{1 \; \rm TeV} \right)^{\! 2} ~.
\label{eq:lambdasd2}
\ee
We assume that $|\lambda_{d}| \ll |\lambda_{s}|$, so that the constraints from kaon measurements are satisfied without imposing an upper limit on $ |\lambda_{s}|$. Combining this assumption with the earlier one made in the discussion of the bounds from $D^0 - \overline {D^0}$, we will neglect the couplings of $\xi$ to first-generation quarks.

Even when the leptoquark does not couple to first-generation quarks, there are nontrivial constraints on the $\lambda_{b} \, \lambda_{s}^*$ product of couplings. These arise from comparing the measurements of the $B_s - \overline {B_s}$ meson mixing with the computation \cite{Becirevic:2015asa}  of the box diagrams with $\xi$ and $\ell^3$ in the loop (replace the $s$ and $\bar d$ fields by $b$ and $\bar s$ in Figure~\ref{fig:KKbar}), which gives 
\be
| \lambda_{b} \, \lambda_{s} | \lesssim 3.6 \times 10^{-2}  \,  \frac{ M_\xi }{1 \; \rm TeV}  ~.
\label{eq:lambdabs}
\ee
Here we used the fit result from \cite{UTfit:2022hsi} for the $B_s$ mass difference: $\Delta m(B_s)/ \Delta m(B_s)_{\rm SM} = 1.14 \pm 0.08$.

Other constraints on $\lambda_{b} \, \lambda_{s}^*$ are set by the measurement of the $B^+ \rightarrow K^+ \, \nu \, \overline \nu$ branching fraction ($B_{K^+}$)  \cite{Belle-II:2023esi}, and by the  existing limit on the 
$B^{ 0} \rightarrow K^{\star 0} \, \nu \, \overline \nu$ branching fraction ($B_{K^{*0}}$) \cite{Belle:2017oht}, which get contributions from the tree-level $b\to s \, \nu_\tau \, \overline \nu_\tau$  transition \cite{Becirevic:2015asa, Browder:2021hbl} (as in Figure \ref{fig:KpinunuDiagram} with appropriate quark replacements). 
The recent measurement $B_{K^+} = (23 \pm 7) \times 10^{-6}$ performed by Belle II \cite{Belle-II:2023esi} is $2.7\sigma$ higher than the SM prediction.
Thus, one can derive both an upper limit and a lower limit on $| \lambda_{b} \, \lambda_{s} | $ at the 95\% CL. However, the lower limit needs further experimental confirmation of potential new physics contributions to $B^+ \rightarrow K^+ \, \nu \, \overline \nu$, and thus we will not consider it as a constraint on this model.\footnote{Reconciling the measurement for $B_{K^+}$ and the experimental limit on $B_{K^{*0}}$ may eventually require new physics coupling to right-handed $b$ and $s$ quarks, generating a right-handed vector interaction, see {\it e.g.}~\cite{Bause:2023mfe}. We will not explore here the range of parameters in our model capable of reconciling these experimental results.}

Using input parameters from \cite{Bause:2023mfe}, we find that the upper  limit imposed by the $B_{K^{*0}}$ limit is stronger than the one from the $B_{K^+}$ measurement, and is given by 
\be
| \lambda_{b} \, \lambda_{s} | \lesssim 3.8 \times 10^{-2}  \, \left(  \frac{ M_\xi }{1 \; \rm TeV} \right)^{\! 2} ~.
\label{eq:lambdabsTree}
\ee
Due to the different dependence on $M_\xi$, this constraint is weaker than (\ref{eq:lambdabs}) for $M_\xi > 950$ GeV.
Given that we assume small $U(1)_\f$ breaking, Yukawa couplings of $\xi$ to $b$ or $s$ quarks are expected to satisfy 
$|\lambda_{b}|, |\lambda_{s}| \lesssim O(0.2)$, so the constraints (\ref{eq:lambdabs}) and (\ref{eq:lambdabsTree}) are rather mild. 


\subsection{Decays of the antileptons} 

For $m_\f \ll M_\xi$ we can integrate out the leptoquark from (\ref{eq:FconservingMass}) and (\ref{eq:FbreakingMass}) 
to obtain the following 4-fermion interactions that violate $\f$ number but preserve lepton number:
\be
\frac{1}{M_\xi^2} \left[ (\lambda_t \overline{t}^c_{ _R} + \lambda_c \overline{c}^c_{ _R} )\, \f_{ _R} \rule{0mm}{4mm} \right]\left[  ( \lambda_b\Bar{b}_{ _R} +  \lambda_s\Bar{s}_{ _R} ) \, \ell_L^3  \rule{0mm}{4mm} \right] ~.
\label{eq:dim6}
\ee
The Lorentz indices here are contracted inside the square brackets, while the $SU(3)\times SU(2)_W$ indices are contracted between the two square brackets. These operators induce 3-body decays of the VALs. There are four final states that involve a charm antiquark,  with partial widths given at leading order by
\bear
&&  \hspace*{-1.2cm}
\Gamma(\f^- \to \overline{c} \,  s \,  \overline \nu_\tau) \simeq \Gamma(\f^0 \to \overline{c} \,  s \,  \tau^+)  \simeq   |\lambda_c \,  \lambda_s|^2 \, \Gamma_0 ~,
\nonumber \\ [-1mm]
&& 
\label{eq:cswidths}
\\ [-1mm]
&&  \hspace*{-1.2cm}
  \Gamma(\f^- \to \overline{c} \,  b  \,  \overline \nu_\tau) \simeq \Gamma(\f^0 \to \overline{c} \,  b \, \tau^+)  \simeq  |\lambda_c  \,  \lambda_b|^2 \, \Gamma_0  ~,
\nonumber 
\eear
where small corrections due to the mass-splitting between the two components of $\f$ are neglected, and we defined
\be
\Gamma_0 = \frac{m_{ _\f}^5}{2048 \, \pi^3 \, M_\xi^4}   ~~.
\ee

For $m_\f \gtrsim m_t + m_\tau \approx 175$ GeV, decays that involve a top antiquark are kinematically allowed:
\be
\Gamma(\f^- \!\to \overline{t} \,  s \,  \overline \nu_\tau) \simeq \Gamma(\f^0 \!\to \overline{t} \,  s \, \tau^+) \simeq   |\lambda_t  \,  \lambda_s|^2 \, \Gamma_0  F_1  ~~,
\ee
with the suppression due to the top mass taken into account through a function of fermion masses:
\be
F_1 =1 - 8\frac{m_t^2}{m_\f^2} + O\left(m_t^4/m_\f^4 \, ,  \, m_\tau^2/m_\f^2   \right) ~.
\ee
For $m_\f \gtrsim m_t + m_b + m_\tau \approx 180$ GeV, there are also decays that involve both a top antiquark and a $b$ quark:
\be
  \Gamma(\f^- \!\to \overline{t} \,  b  \,  \overline \nu_\tau) \simeq \Gamma(\f^0 \!\to \overline{t} \,  b \, \tau^+)  \simeq   |\lambda_t  \,  \lambda_b|^2 \, \Gamma_0 F_2 ~~,
\label{eq:tbwidths}
\ee
where 
\be
F_2 \approx F_1 + O\left(m_b^2/m_\f^2\right)  ~.
\ee
If  $|\lambda_c | \ll  |\lambda_t|$, the decay modes of $\f$ that involve its Yukawa coupling to the top quark may dominate even for 
$m_\f < m_t$. In that case, the top quark is off-shell, and the $\f$ VALs have 4-body decays:   $\f^- \!\to W^- \overline{b} \,  s \,  \overline \nu_\tau$ and $\f^0 \!\to W^- \overline{b} \,  b \, \tau^+$.

Let us now consider the limit $|\lambda_t| \ll |\lambda_c|$, which implies that the total widths of $\f^-$ and $\f^0$ are dominated by 
the partial widths (\ref{eq:cswidths}). The ensuing decay length in the rest frame for the $\f$ VALs is 
\begin{align}
c \,  \tau_{\! _\f} =\frac{1.25 \times 10^{-7} \, \text{cm}}{|\lambda_c|^2 ( |\lambda_s|^2 +  |\lambda_b|^2)} \left(\frac{M_\xi}{1 \, \text{TeV}}\right)^{\! 4} \left(\frac{100 \, \text{GeV}}{m_{ _\f}}\right)^{\! 5} ~.
\label{eq:length}
\end{align}
Even after taking into account the typical boost of the $\f$ pair produced at the LHC,  the
 $\f$ VALs decay promptly into visible final states provided $c\, \tau_{\! _\f} \lesssim 10^{-2}$cm.
We focus on this scenario in what follows.

A small mass-splitting between the two components of the $\f$ doublet is induced at one loop. The contributions from the $Z$ boson or photon running in the loop have been computed~\cite{Thomas:1998wy}, and give $\delta m_\f = m_{\f^-}\! - m_{\f^0} > 0$. This mass-splitting is only logarithmically dependent on $m_\f$, and continuously grows from 250 MeV to 350 MeV when $m_\f$ grows from 100 GeV to 1 TeV. This mass-splitting allows a charged-current decay, $\f^{-} \to \f^0 \pi^{-}$, with a width
\begin{align}
\hspace{-0.2cm}
    \Gamma (\f^\pm \! \to \f^0 \pi^\pm) \simeq \frac{G_F^2}{\pi}  \, f_\pi^2 \, (\delta m_\f)^3 
    \left( 1-\frac{m_\pi^2}{ (\delta m_\f)^2}   \right)^{\! 1/2}  ,
\end{align}
where $f_\pi  \approx 130 $ MeV is the pion decay constant, and $G_F$ is the Fermi constant.
The decay length in the rest frame associated with $\f^{-} \to \f^0 \pi^{-}$ is above 0.6 cm, and thus much longer than the one given in (\ref{eq:length}). Consequently, the branching fraction for $\f^{-} \to \f^0 \pi^{-}$ is small enough so we can neglect this process.

\begin{figure}[t]
\centering
\vspace*{-5.5mm}
\hspace*{-1mm} \includegraphics[scale=0.628]{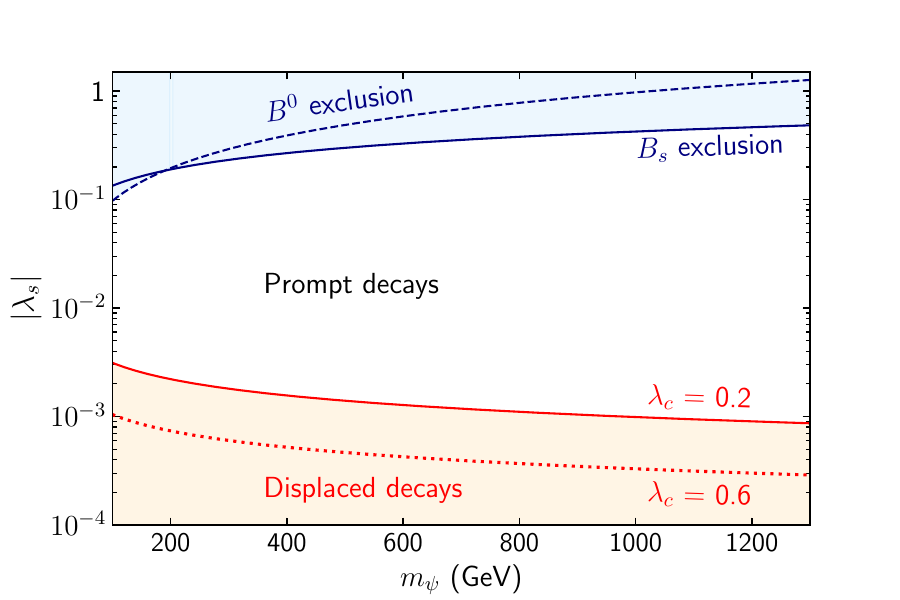}
\vspace*{-4.5mm}
\caption{Constraints on $|\lambda_s|$ as a function of the $\f$ mass, where $\lambda_s$ is the Yukawa coupling of the 
$s$ quark to the $\xi$ leptoquark. Here we fix the $b$ quark coupling $\lambda_b = \lambda_s$.
The unshaded region is allowed and corresponds to prompt $\f$ decays, for $M_\xi = 5 m_\f$. 
The upper shaded (blue) region is ruled out by $\xi$ contributions to $B_s - \overline B_s$ mixing (above the solid blue line) 
and $B^{0} \rightarrow K^{\star 0} \, \nu \, \overline \nu$ (above the dashed blue line). 
The lower shaded (orange) region corresponds to $\f$ displaced vertices (rest-frame  decay length $c\, \tau_{\! _\f} < 10^{-2}$cm) for
a $c$-quark coupling $\lambda_c = 0.2$ and $|\lambda_t/\lambda_c| \ll 1$. 
The edge of that region moves to the  dotted red line when $\lambda_c=0.6$. 
 }
\label{fig:lambdas}
\end{figure}

An implication of the expression for the decay length in (\ref{eq:length}) is that the couplings $\lambda_s$ and $\lambda_b$, although expected to be small due to the $U(1)_\f$ symmetry, cannot both be arbitrarily small if the $\f$ VALs decay promptly in the detector.
In Figure~\ref{fig:lambdas}, the unshaded region corresponds to prompt $\f$ decays ($c\, \tau_{\! _\f} < 10^{-2}$cm) and 
is allowed by the $B$ meson constraints  (\ref{eq:lambdabs}) and (\ref{eq:lambdabsTree}).
The mass ratio is fixed there at $M_\xi/m_\f = 5$, which is near the lower end that allows\footnote{For a smaller mass ratio, higher-order terms in $(M_\xi/m_\f)^{-1}$ would need to be included. The overall impact would be a shorter decay length of $\psi$, and also stronger constraints from flavor processes, so the allowed region of prompt decays in  Figure~\ref{fig:lambdas} would shift towards smaller $|\lambda_s|$ values.} us to  integrate out $\xi$ and 
use the 4-fermion operators (\ref{eq:dim6}). For larger $M_\xi/m_\f \equiv r_m$, all the lines in Figure~\ref{fig:lambdas} move up by the following factors: 
$\sqrt{r_m/5}$ for the upper solid line, $r_m/5$ for the dashed line, and $(r_m/5)^2$ for the lower solid and dotted lines.

The Yukawa coupling of the $b$ quark to the $\xi$ leptoquark is fixed in Figure~\ref{fig:lambdas} at  $|\lambda_b| = |\lambda_s|$.
For a different $\lambda_b$, the dashed and upper solid lines scale as $|\lambda_s/\lambda_b|^{1/2}$, while the dotted and lower solid lines 
scale as $\sqrt{2}/(1+|\lambda_b/\lambda_s|^2)^{1/2}$.
If instead of the $|\lambda_t/\lambda_c| \ll 1$ limit assumed in Figure~\ref{fig:lambdas}, 
we consider the opposite limit of dominant  $\lambda_t$, then the region of prompt decays is reduced. This is because these decay widths decrease due to the additional phase-space suppression for final states with a top quark.

For $m_{ _\f} < m_t$, the top quark can decay through an off-shell $\xi$ into 3-body final states,
such as $t \to  \overline{\f} {}^{\, 0}  \, b \, \tau^+$ and $t \to \f^+ \, b \, \overline \nu_\tau$. The branching fractions of these processes are 
proportional to $|\lambda_t \lambda_b|^2$, and are highly phase-space suppressed. Nevertheless, it is worth searching for peculiar cascade decays of the top quark, for example $t \to  \overline{\f} {}^{\, 0}  \, b \, \tau^+ \to  (c \, \bar b \, \tau^-)  \, b \, \tau^+ $.

\medskip\medskip

\section{Collider 
phenomenology}
\label{sec:pheno}

The global $U(1)_\f$ symmetry of (\ref{eq:Fconserving}) implies that direct production of $\psi$ VALs at colliders can proceed only in pairs,
through electroweak processes. Although the explicit $U(1)_\f$ breaking allows in principle single leptoquark production, followed by $\xi \to \f \, c$ or $\xi \to \f\,t$, the small couplings in (\ref{eq:FbreakingMass}) and the need for a $\tau$ in the initial state effectively suppress this channel.
Effects of a $\xi_u$ exchanged in the $t$-channel \cite{Haisch:2022lkt} lead to $s \bar s \to \tau^+\tau^-$ and $b \bar b \to \tau^+\tau^-$ partonic processes (the latter is further PDF suppressed), but these are suppressed\footnote{Recall that $|\lambda_s|, |\lambda_b| \ll 1$ because it is natural to impose that the global $U(1)_\f$ symmetry is approximately preserved.
If that assumption is relaxed, then traditional leptoquark searches \cite{Diaz:2017lit,Dorsner:2016wpm} may dominate.} by $|\lambda_s|^4$ or $|\lambda_b|^4$.

It is also possible to pair produce the leptoquarks, with one of them decaying to a $\psi$ and the other one decaying through  the $U(1)_\f$ breaking 
couplings (\ref{eq:FbreakingMass}). This would lead to a final state with a single $\psi$, but we will neglect it in what follows as the associated branching fractions are likely to be very small.

In this Section, we will study pair production of the new particles, followed by decays via couplings involving SM quarks of either the second or third generation. 
We will not address here the case where the couplings to the second- and third-generation quarks are comparable, which would imply 
sizable branching fractions for final states involving $t$ and $s$ quarks, or $b$ and $c$ quarks. 

\subsection{Pair production of fermion doublet at the LHC}

The $ \f^{0}$ and $\f^{-}$ fermions, which carry lepton number $L= -1$, have antiparticles 
$\overline \f {}^{\, 0}$ and $\f^{+}$ fermions that carry lepton number $L= +1$.
There are four main mechanisms for $\f$  production at hadron colliders:
\bear
& u \, \overline d    \to \, W^{+*}  \to \; \f^+  \, \f^0   ~~,
\nonumber \\ [2mm]
& d \, \overline u    \to \, W^{-*}  \to \; \f^-  \, \overline{\f} {}^{\, 0}   ~~,
\nonumber \\ [-1.5mm]
&   
\label{eq:partonic}
 \\ [-1.5mm]
& q \,\overline q  \to  \, Z^{*}   \to  \; \f^0 \, \overline{\f} {}^{\, 0}   ~~,
\nonumber \\ [2mm]
& q \,\overline q    \to \,  \gamma^*, Z^{*}   \to \; \f^+ \f^-   ~~.
\nonumber
\eear
Each of the above $s$-channel electroweak boson is off-shell, and only the dominant partonic processes are shown here.

To compute the cross sections for production of each of the four fermion pairs in proton-proton collisions, we generated model files using
FeynRules~\cite{Alloul:2013bka} and run simulations using MadGraph\_aMC@NLO~\cite{Alwall:2014hca}. The PDF set used here is NNPDF4.0 NLO \cite{NNPDF:2021njg}. The renormalization and factorization scales were set dynamically by MadGraph\_aMC@NLO.
  
\begin{figure}[t]
\centering  \hspace*{-2mm}
\includegraphics[scale=0.73]{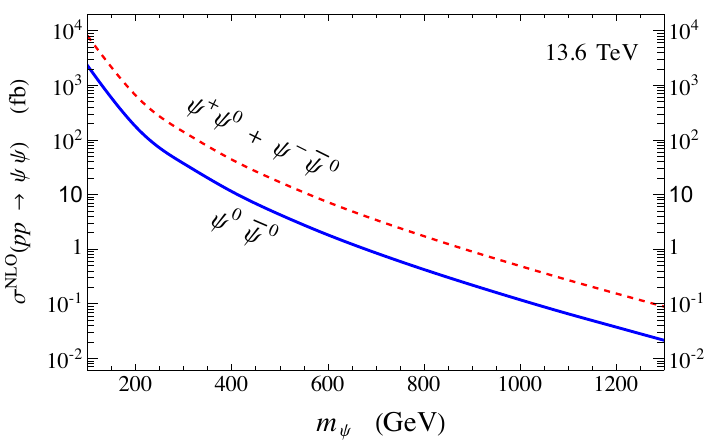}
\caption{NLO cross sections for pair production of a weak-doublet fermion  through $s$-channel off-shell electroweak bosons at the 13.6 TeV LHC. 
The sum of the cross sections for $pp \to W^{+*}\to \psi^+ \psi^{0}$ and $pp \to W^{-*} \to \psi^-  \overline \psi  {}^{\, 0}$  is given by the dashed red line.
The cross section for $pp \to Z^* \to \psi^0 \overline \psi {}^{\, 0}$ is given by the solid blue line, and is almost equal to the cross section for $\psi^+ \psi^-$ production (see Figure \ref{fig:xsratio}).}
\label{fig:xsection}
\end{figure}

In Figure~\ref{fig:xsection} we show the $\psi$ pair production cross sections, at the next-to-leading order (NLO) in $\alpha_s$, 
as a function of the $\psi$ mass, for $pp$ collisions at $\sqrt{s} = 13.6$ TeV, which is the center of mass energy of the current Run 3 at the LHC.
The sum of the cross sections  for the $pp \to W^{+*}\to \psi^+ \psi^0$ and $pp \to W^{-*} \to \psi^- \overline \psi {}^{\, 0} $ processes\footnote{For brevity, processes such as $pp \to W^{-*}  + X \to \psi^- \psi^0 +X$, where $X$ represents hadronic activity not related to the partonic processes (\ref{eq:partonic}), are denoted here without the $X$. }
is shown as the dashed red line, while the cross section for $pp \to Z^* \to \psi^0 \overline \psi {}^{\, 0}$ is shown as the solid blue line.

\begin{figure}[t]
  \vspace*{1.5mm}
\centering  \hspace*{-4mm}
\includegraphics[scale=0.68]{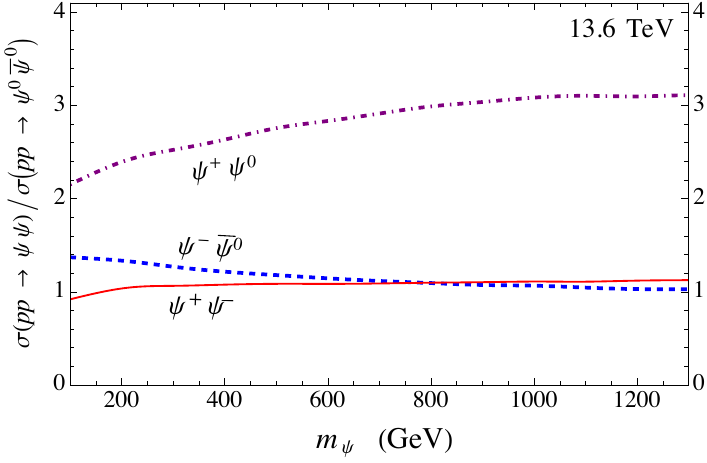}
\caption{Ratios of cross sections for $\f$ pair production ($\psi^+ \psi^0$, $\psi^- \overline \psi {}^{\, 0}$, $\psi^+ \psi^-$) to the $pp \to Z^* \to \psi^0 \overline \psi {}^{\, 0}$ cross section at a center-of-mass energy of 13.6 TeV.}
\label{fig:xsratio}
\end{figure}

The cross sections for the $pp \to \gamma^*, Z^* \to \psi^+ \psi^-$ and $pp \to Z^* \to \psi^0 \overline \psi {}^{\, 0}$ processes differ by less than 10\% for $m_\psi$ in the 0.1--1 TeV range. Furthermore, the cross sections for  $pp \to W^{-*} \to \psi^- \psi^0$ and $pp \to Z^* \to \psi^0 \overline \psi {}^{\, 0}$ differ by less than 15\% for $m_\psi > 0.6$ TeV.  These differences are accidentally small due to the interplay between the PDFs and the electroweak couplings of the quarks and $\psi$. In Figure \ref{fig:xsratio} we show the following ratios of cross sections: 
$\sigma(pp \to \gamma^*, Z^* \to \psi^+ \psi^-)/ \sigma(pp \to \psi^0 \overline \psi {}^{\, 0})$ as a solid red line,
$\sigma(pp \to W^{-*} \to \psi^- \psi^0)/ \sigma(pp \to \psi^0 \overline \psi {}^{\, 0})$ as a dashed blue line, and 
$\sigma(pp \to W^{+*}\to \psi^+ \overline \psi {}^{\, 0})/ \sigma(pp \to \psi^0 \overline \psi {}^{\, 0})$ as a purple dash-dotted line.  


Note that the ratio of the cross sections for the processes mediated by $W^{+}$ and $W^{-}$ grows from about 1.7 to 2.9 when $m_\psi$ is increased from 0.1 to 1 TeV because the ratio of the $u$ and $d$ PDFs is approximately 2 at low parton momentum fraction ($x$) where the valence quarks of the protons are counted.

The effective couplings (\ref{eq:dim6}) allow four decay modes for each of the $\psi$ fermions. To reduce the number of final states,
we now adopt a flavor ansatz for the coupling of the VALs and leptoquarks to quarks:  
$\xi$ and $\f$ couple predominantly to one generation of SM quarks at a time. 

\subsection{Couplings to second-generation quarks}  
\label{sec:2gen}

In this section, we assume that the couplings to second-generation quarks  dominate, {\it i.e.},  $|\lambda_b|^2 \ll |\lambda_s|^2$ and  $|\lambda_t|^2 \ll |\lambda_c|^2$, and study the collider signals of the $\f$ and $\xi$ particles. 

\subsubsection{Signals of the $\f$ antileptons with taus and jets}

\begin{figure}[t]
    \centering
\includegraphics[scale=0.96]{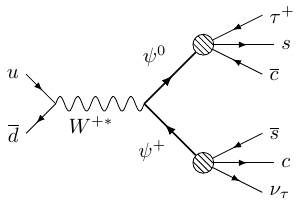}    \\
\vspace*{3mm}
\includegraphics[scale=0.96]{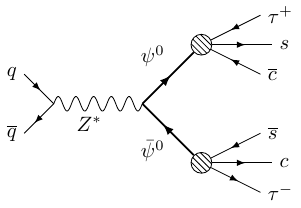}
    \caption{Representative diagrams for the production and decay of the VALs. The blobs represent the
    effective interactions mediated by a heavy $\xi$ leptoquark. 
    The charged-current process leads to a $ \tau^+ + 4j + \slash \!\! \!\! E $ signal, while the process mediated by an off-shell $Z$ leads to a $ \tau^+\tau^- + 4j$ signal.
}
    \label{fig:diagrams-prod-decay}
\end{figure}

The main decay modes of the VALs in the case of dominant couplings to second-generation quarks are 
\bear
&   \f^0 \to \overline{c} \,  s \,  \tau^+  \;\;\;\;   ,  \;\; \;\;   \overline{\f} {}^{\, 0}  \to c \, \overline{s}  \,  \tau^-    ~~,
\nonumber \\ [-1.5mm]
&
 \\ [-1.5mm]
&  \f^- \to \overline{c} \,  s \,  \overline \nu_\tau    \;\;\;\;   ,  \;\; \;\;     \f^+ \to c \, \overline{s} \,  \nu_\tau  ~~.
\nonumber 
\eear 
The process of the largest cross section involving $\f$ VALs at proton-proton colliders is then 
\be
p \, p  \, \to W^{\pm*}    \to \; \f^\pm  \overset{\textbf{\fontsize{0.1pt}{1pt}\selectfont \,(--)}}{\f}  \!^{\, 0}   
\to \;  (\tau^\pm \, jj) \,  (  \, \overset{\textbf{\fontsize{1pt}{1pt}\selectfont  (--)}}{\nu} jj) ~~,
\label{eq:Zpmcs}
\ee
as can be seen in the first diagram of Figure~\ref{fig:diagrams-prod-decay}.
The signal  is thus $ \tau + 4j + \slash \!\! \!\! E $, where $j$ stands for any hadronic jet that is not $b$-tagged. Two of the four jets can in principle be tagged as $c$-jets, but that would reduce the signal rate because currently the efficiency of $c$-tagging is relatively low. The missing transverse energy ($\slash \!\! \!\! E $) due to the neutrino is roughly $m_\psi/3$, so it provides limited means of rejecting the background. 

An irreducible background to the $ \tau + 4j + \slash \!\! \!\! E $ signal is SM $W+4j$ production followed by $W \to \tau \nu$, which has an inclusive cross section\footnote{The cross sections quoted in \cite{Anger:2017nkq} are smaller because they are computed in the presence of additional cuts. We thank Stefan H\"{o}che for providing the $W+4j$ cross sections computed with Sherpa \cite{Sherpa:2019gpd} in the presence of only the jet $p_T > 30$ GeV and $\eta < 5$ cuts.} times branching fraction of the order of 70 pb \cite{Anger:2017nkq} when the transverse momentum of each jet is above 30 GeV.
Since this is at least an order of magnitude larger than the signal even for $m_\psi$ near 100 GeV, and there are additional challenging backgrounds (such as pure QCD in the case of hadronic tau decays), these charged-current channels can be probed only if the experimental collaborations would perform advanced, dedicated searches.

Another channel is the neutral-current process
\be
p \, p  \to  \, Z^{*}   \to  \; \f^0 \, \overline{\f}  {}^{\, 0}  \to \;   (  \tau^+ j j ) (  \tau^- j j )    ~~,
\label{eq:Z00cs}
\ee
which is shown in the second diagram of Figure~\ref{fig:diagrams-prod-decay}.
This $ \tau^+ \tau^- + 4j$ signal has a smaller cross section  (see Figures~\ref{fig:xsection}), but the backgrounds are also significantly smaller. Most notable
backgrounds include SM production of $Z+4j$ followed by $Z \to \tau^+ \tau^-$,  or $W^+W^-+4j$  followed by each $W$ boson decaying to $\tau \nu$. The small cross sections of these backgrounds, and the additional opportunities at rejecting the background, make 
$ \tau^+ \tau^- + 4j$ the most sensitive channel.

A CMS search for a $ \tau^+ \tau^- + 4b$ signal \cite{CMS:2022cpe}, arising from pair production of a vectorlike lepton doublet that decays through an off-shell vector leptoquark \cite{DiLuzio:2017vat,Bordone:2017bld, DiLuzio:2018zxy,Greljo:2018tuh}, has set a lower mass limit of 500 GeV. That CMS search, based on Run 2 data,  relied on at least three $b$-tags to reduce the backgrounds. Given that our $ \tau^+ \tau^- + 4j$ signal does not include $b$ jets, the highest $m_\psi$ that a dedicated search with Run 2 data can explore is substantially below 500 GeV.

The remaining $\psi$ pair production channel,
\be
p \, p  \to  \, \gamma^*, Z^{*}   \to \; \f^+ \f^-  \to \; ( \nu j j ) (\overline \nu j  j)  ~,
\ee
leads to a difficult $4j + \slash \!\! \!\! E $ signal. Searches at the LHC in similar channels are 
sensitive to cross sections  \cite{CMS:2019zmd}  larger by an order of magnitude compared to the one for $\f^+ \f^-$ given in Figures~\ref{fig:xsratio} and Figures~\ref{fig:xsection}. Nevertheless, future dedicated searches for the VALs may combine this channel with  $ \tau^+ \tau^- + 4j$ and $ \tau + 4j + \slash \!\! \!\! E $ to increase the sensitivity. 

We will not attempt to predict the range of $m_\psi$ that the ATLAS and CMS collaborations can explore, but we can safely conclude that 
the mass reach in the $\psi$ pair production channels is much smaller than in the case of a vectorlike lepton doublet that decays into two SM particles via mass mixing with the tau (the current lower mass limit in that case is 1.05 TeV \cite{CMS:2022nty}). 

The cross section for pair production of VALs at a muon collider or an $e^+e^-$ collider is large enough \cite{Bhattiprolu:2023yxa, Baspehlivan:2022qet},  and the backgrounds are much smaller than at hadron colliders. Therefore, a lepton collider with $\sqrt{s} \ge 2 m_\psi$ would allow a thorough investigation of the VAL properties.

\subsubsection{Signals of the $\xi$ leptoquarks with taus and jets}

\begin{figure}[t!]
    \centering
\includegraphics[scale=1]{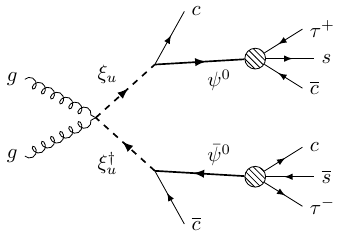}
    \caption{Representative diagram for pair production of the $\xi_u$ leptoquark followed by cascade decays, leading to a $\tau^-\tau^+ + 6j$ signal at hadron colliders. }
    \label{fig:LQdecay}
\end{figure}

Given that the breaking of the global $U(1)_\f$ symmetry is small, and that the couplings to second-generation quarks are assumed to be dominant in this Section, the largest interaction of the leptoquark doublet $\xi$ is provided by the 
$\lambda_{c} \, \xi^\dagger  \overline{c}^{\,  c}_{ _R}   \f_{\! _R}$ term in  
(\ref{eq:FconservingMass}). 
This implies that the dominant decay modes for the $\xi_u$ and $\xi_d$ leptoquarks are
\be
\xi_u \to \, c \, \f^0    \;\;\; , \;\;\;    \xi_d \to \, c \, \f^-  ~~,
\ee
for $M_\xi > m_\psi$. Each $\f$ fermion produced in a $\xi$ decay subsequently undergoes a 3-body decay into SM fermions, mediated by an off-shell $\xi$. As a result, leptoquark pair production is followed by the following cascade decays
\bear
& \hspace*{-7mm}
p \, p \to \xi_u \,  \xi_u^\dagger  \to \, (c \, \f^0 ) \, (\overline c \; \overline\f^0)   \to \, ( \tau^+ jjj) \, (\tau^- jjj)   ~,
\nonumber \\ [-1.5mm]
&
 \\ [-1.5mm]
& \hspace*{-7mm}
p \, p \to \xi_d \,  \xi_d^\dagger  \to \, (c \, \f^- ) \, (\overline c \, \f^+)   \to \, (\overline\nu_\tau jjj) \, (\nu_\tau jjj)  ~.
\nonumber 
\eear
Thus, pair production of the leptoquark of electric charge +2/3 ($\xi_u$) leads to a $\tau^-\tau^+ + 6j$ signal with a topology shown in Figure \ref{fig:LQdecay}.

Existing searches for  leptoquarks coupling to a tau and a quark assume a $(\tau b)(\tau b)$  or  $(\tau t)(\tau t)$  signal, 
and set lower mass limits of 1.19 TeV \cite{ATLAS:2021jyv} and 1.43 TeV \cite{ATLAS:2021oiz, CMS:2022nty}, respectively.
Compared to the existing $(\tau b)(\tau b)$ search, our $\tau^-\tau^+ + 6j$ signal does not include $b$ jets and has much softer taus and jets due to the larger number of final-state particles. Consequently, the backgrounds are much larger, so we expect that the mass reach of even a dedicated search for $\xi_u$ with the Run 2 data would be below 1 TeV. 

Pair production of the leptoquark of electric charge $-1/3$ ($\xi_d$) leads to a $6j + \slash \!\! \!\! E $ signal.
Current leptoquark searches in the jets + $\slash \!\! \!\! E $ channel \cite{CMS:2019ybf} assume a $(\nu j)(\nu j)$ signal, and set a lower limit 
mass limit of 1.14 TeV. As our signal  is plagued by much larger backgrounds, due to larger jet transverse momenta and more missing transverse energy, we again expect that $M_\xi$ can be substantially below 1 TeV.
Thus, the $\xi$ leptoquarks are clearly more elusive than the leptoquarks searched so far in LHC experiments. 

\subsection{Couplings to third-generation quarks}  
\label{sec:3gen}

Let us now consider the case where the couplings to third-generation quarks  dominate, {\it i.e.}, the Yukawa couplings in Eqs.~(\ref{eq:FconservingMass}) and (\ref{eq:FbreakingMass}) satisfy
 $|\lambda_s|^2 \ll |\lambda_b|^2$ and  $|\lambda_c|^2 \ll |\lambda_t|^2$.    
 As this scenario leads to signals involving $t$ and $b$ quarks, the $\psi$ and $\xi$ particles are less elusive than in the case of 
 couplings to second-generation quarks discussed in Section~\ref{sec:2gen}.
 
\subsubsection{Signals of the $\f$ antileptons with top quarks}
\label{sec:psit}

\begin{figure}[t!]
    \centering
\includegraphics[scale=0.96]{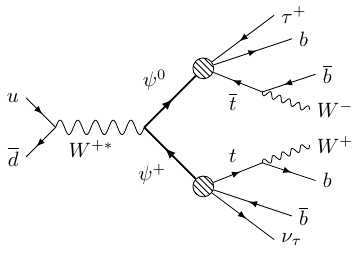}  
    \caption{Pair production of VALs through an off-shell $W^+$, followed by decays induced by couplings to third-generation quarks. 
    This process leads to a $ \tau^+ W^+W^- \! + 4b + \slash \!\! \!\! E $ signal.}
    \label{fig:diagrams-prod-decay-tb}
\end{figure}

If the VALs are sufficiently heavy, $m_\psi > m_t + m_b + m_\tau \approx 180$ GeV, then their main decay modes are:  
\be
 \f^0 \to \overline{t} \,  b \,  \tau^+  \;\;\;\;   ,  \;\; \;\;   \f^- \to \overline{t} \,  b \,  \overline \nu_\tau   ~~.
 \label{eq:psitb}
\ee
In this case the main signal at proton-proton colliders is $ \tau^\pm W^+W^- \! + 4b + \slash \!\! \!\! E $, and arises from the diagram 
with an off-shell $W^+$ shown in Figure~\ref{fig:diagrams-prod-decay-tb} and the analogous diagram with an off-shell $W^-$.

Tagging three or four of the $b$-jets  substantially reduces several backgrounds ({\it e.g.}, $t\bar t$ plus a $W$ or a $Z$, and $WZb\bar b$) without significantly depleting the signal. 
For a $b$-tagging efficiency $\epsilon_b$ in the $0.8-0.9$ range \cite{CMS:2017wtu,Bols:2020bkb}, the combined efficiency of at least three $b$-tags, $\epsilon_{3b} = \epsilon_b^3 (4 - 3 \epsilon_b)$, is in the  $82\%-95\%$ range.  

Nevertheless, irreducible backgrounds with four $b$ jets exist in the SM, including $t\bar t h^0$ or $WW+4b$. 
The presence in the final state of two muons or electrons of same charge, due to  leptonic decays of the $\tau$ or $W$ bosons, suppresses these backgrounds, but at the cost of also reducing the signal.
The combined branching fractions for the $ \tau^\pm W^+W^- \! + 4b + \slash \!\! \!\! E $ final state to include 
two same-sign charged leptons ($\ell^\pm\ell^\pm$ where $\ell$ stands for an electron or a muon), or three charged leptons ($\ell^\pm\ell^+\ell^-$), are
\bear
&& B_{\ell^\pm\ell^\pm}  = 2 B_{\tau\ell} \, B(W\to \ell \bar \nu) \left(2 + B_{\tau\ell} \rule{0mm}{3.7mm}  \right)  \approx  8.3\% ~,
\nonumber \\ [-2mm]
\label{eq:Bll}
\\ [-2.5mm]
&& B_{3\ell}  = 2 B_{\tau \ell} \, B(W\to \ell \bar \nu)^2 \left(2 + B_{\tau \ell} \rule{0mm}{3.7mm}  \right)^{\! 2}   \approx  2.0\% ~,
\nonumber 
\eear
where $B_{\tau  \ell}$ is the average of the $\tau$ branching fractions into $\mu \nu \bar \nu$ and $e \nu \bar \nu$,
and lepton universality is taken as a good approximation in the $W$ branching fractions \cite{ParticleDataGroup:2022pth}. 

Even these $\ell^\pm\ell^\pm + (\ge \! 3) b$ and $3\ell + (\ge \! 3) b$  signals have SM backgrounds, especially from 
$t \,\bar t \, t \, \bar t$ and $t \, t \, \bar t$ productions.
There are also important backgrounds due to leptons produced in meson decays or in photon conversions, and due to the misidentification of the lepton charge or of the $b$ tags.

Despite all these complicated backgrounds, we could have a rough estimate for the Run 3 reach in $\psi$ pair production by comparing 
to the $pp \to t \, \bar t \, t \, \bar t$ process,  which has recently been observed by the ATLAS and CMS collaborations \cite{ATLAS:2020hpj,CMS:2023ftu}. 
With 138--139 fb$^{-1}$ of data at $\sqrt{s} =13$ TeV, the measured $ t \, \bar t \, t \, \bar t$ cross section ($\sigma_{4t}$) has a central value of about 18 fb for CMS and 24 fb for ATLAS, and are within $1\sigma$ of each other and consistent with the SM prediction \cite{vanBeekveld:2022hty}.  
The combined leptonic branching fractions for $\psi$ signals, given in (\ref{eq:Bll}), are smaller than the analogous ones for $t \, \bar t \, t \, \bar t$ 
($B_{\ell^\pm\ell^\pm} \approx 11\%$, $B_{3\ell} \approx 4.3\%$).
Furthermore, the backgrounds are larger at $\sqrt{s} =13.6$ TeV compared to 13 TeV,
so even with a larger integrated luminosity of 300 fb$^{-1}$ the cross section sensitivity to $\psi$ pair production cannot be smaller than roughly 20 fb ({\it i.e.}, about the measured $\sigma_{4t}$ in Run 2). This corresponds to a mass reach $m_\psi \lesssim 500$ GeV, as can be seen in Figure~\ref{fig:xsection}.

The reach in $m_\psi$ can be increased by adding the  $pp \to Z^* \to  \psi^0 \overline \psi^0 \to  \tau^+ \tau^- W^+W^- \! + 4b $ process, which has a smaller production cross section (by a factor of about 4, see Figures  \ref{fig:xsection} and \ref{fig:xsratio}) but larger leptonic branching fractions 
($B_{\ell^\pm\ell^\pm} \approx 17\%$, $B_{3\ell} \approx 10\%$).
This process also leads to final states with hadronic $\tau$ decays that would further increase the mass reach. Even though the $pp \to \psi^+\psi^- \to t \, \bar t \, b \, \bar b + \slash \!\! \!\! E$ process does not produce two same-sign leptons, it can be useful due to the missing transverse energy, which suppresses the $ t \, \bar t \, b \, \bar b$ background. 

Combining all these channels and optimizing the event selections would improve the sensitivity, so that  the largest $m_\psi$
that  can  be probed in Run 3 is likely in the 600--700 GeV range. Note that this is again significantly below the lower mass limit of 1.05 TeV set by Run 2 searches \cite{CMS:2022nty} for vectorlike lepton doublets undergoing standard decays. 

\subsubsection{Signals of the $\xi$ leptoquarks with top quarks}

\begin{figure}[t!]
    \centering
\includegraphics[scale=1]{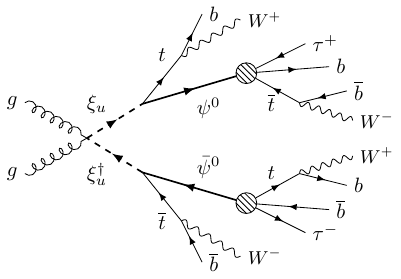}
    \caption{Representative diagram for pair production of the $\xi_u$ leptoquark followed by cascade decays via the $\psi$ VALs, leading to a $\tau^-\tau^+b\bar b + 4t$ signal at hadron colliders. }
    \label{fig:LQdecaytb}
\end{figure}

Consider next the $\xi$ leptoquark signals in the case of $\psi$ decays involving third-generation quarks shown in (\ref{eq:psitb}).
Pair production of the charge $2/3$ leptoquark followed by cascade decays through the $\psi$ VALs 
leads, as shown in the diagram from Figure~\ref{fig:LQdecaytb}, to a $\tau^+\tau^-b\bar b + 4t$ final state.
Thus, the signal is $\tau^+\tau^- + 6b + 4W$, and includes a large number  (up to 6) of charged leptons in addition to the six 
$b$ jets.
These provide various ways of eliminating the backgrounds. On the other hand, requiring many leptons implies small combined branching fractions which would reduce the signal too much. Moreover, the large multiplicity of particles in the final state implies that most of them are relatively soft. Following the discussion in Section~\ref{sec:psit}, we expect that the reach in $\xi$ pair production cross section is of the same order of magnitude as for $4t$, so around 20 fb. This corresponds to $M_\xi \lesssim 0.9$ TeV \cite{Diaz:2017lit, Mandal:2015lca, Kramer:2004df}, which is substantially below the Run 2 lower mass limits (in the 1.2--1.4 TeV range \cite{ATLAS:2021jyv, ATLAS:2021oiz, CMS:2022nty,ATLAS:2021yij}) for leptoquarks that have 2-body decays into SM particles.

\subsubsection{Signals for $m_\psi$ in the $100 - 180$ GeV range}

Finally, if $m_\psi$ is in the $\sim 100 - 180$ GeV range, then the $\psi$ decays shown in (\ref{eq:psitb}), which involve a top quark, are kinematically closed. In that case, $\psi^0$ and $\psi^-$ may undergo 4-body decays  through an off-shell top quark
($\psi^0 \to W^- b \, \bar b \, \tau^+$ and $\psi^- \to W^- b  \,  \bar b  \, \nu_\tau$). If these are the main decay modes, then the additional phase-space suppression is likely to make the decay length long enough so that a highly-ionizing charged track due to $\psi^\pm$ can be observed.
Another possibility is that $|\lambda_c|$ is large enough (while satisfying $|\lambda_c|^2 \ll |\lambda_t|^2$ so that the couplings to third-generation quarks dominate), such that the 3-body decays involving a charm quark,  
\be
\psi^0 \to \bar c  \,   b  \,  \tau^+ \;\; , \;\;  \psi^- \to \bar c  \,  b  \,  \overline \nu_\tau   ~~,
\ee
have larger widths than the 4-body decays, and are prompt.  We focus on this case in the remainder of this Section.

Pair production of $\psi$ then includes the following signals at the LHC:
\bear
& p \, p  \, \to W^{\pm*}    \to \; \f^\pm  \, \overset{\textbf{\fontsize{0.1pt}{1pt}\selectfont \,(--)}}{\f}  \!^{\, 0}    \to \;   \tau^+ b\bar b j j 
+   \slash \!\! \!\! E    ~~,
\nonumber \\ [-1.5mm]
&
 \\ [-1.5mm]
& \hspace*{-7mm}
p \, p  \, \to Z^*  \to \; \f^0 \,  \overline\f {}^0 \to \tau^+\tau^- b\bar b j j    ~~.
\nonumber 
\eear
Although the presence of $b$ jets avoids some of the backgrounds discussed in Section~\ref{sec:2gen}, these signals remain challenging.   
The above charge-current process has very large irreducible backgrounds from $t\bar t$ and $Wbbjj$ productions. Even the 
neutral-current process may be hard to observe due to $t\bar t jj$ and $W$ + jets backgrounds.

It may be that the discovery modes for the VALs in this case are via the decays of the leptoquarks. Note that the latter decays involve top quarks
even when $\psi$ is below the $m_t$ threshold. 
The most interesting process for $m_\psi$  below the top threshold is
\be
p \, p \to \xi_u \,  \xi_u^\dagger  \to \, (t \, \f^0 ) \, (\overline t \; \overline\f {}^0)   \to \, ( \tau^+ t \, b \, j) \, (\tau^-  \bar t \, \bar b \, j)   ~.
\ee
The signal includes four $b$ jets and up to four charged leptons, so if $M_\xi$ is roughly in the $0.5 - 1$ TeV range, then it is possible that a dedicated search in this channel would discover both $\xi_u$ and $\f^0$ in Run 3 of the LHC.

\bigskip

\section{Conclusions}
\label{sec:conc}

The extension of the SM proposed here includes two new fields, which are weak doublets: a vectorlike fermion $\f$ transforming as $(1,2,-1/2)$ under the SM gauge group, and a scalar leptoquark $\xi$ transforming as $(3,2,+1/6)$. Gauge invariance allows a single type of renormalizable interaction involving both these fields, namely a Yukawa coupling to the SM up-type weak-singlet quarks, as displayed in  (\ref{eq:FconservingMass}). 
Likewise, there is a single type of interaction involving only $\xi$ and SM fields: a Yukawa interaction of a down-type weak-singlet quark with a SM lepton doublet (assumed here to be the third-generation one) and $\xi$, of the form shown in (\ref{eq:FbreakingMass}).
Both these types of Yukawa interactions conserve lepton number provided both $\f$  and $\xi$ carry lepton number $L = -1$.
Since the $\f$ doublet has the gauge charges of an often studied vectorlike lepton, but opposite lepton number, we use the term vectorlike antilepton, or VAL. 

Interactions  (\ref{eq:FconservingMass}) preserve a global $U(1)_\f$ symmetry under which $\f$ and $\xi$ carry the same charge, so it is natural to expect that the interactions (\ref{eq:FbreakingMass}), which break that symmetry, have smaller couplings. This coupling hierarchy implies that the  $\xi$ scalars elude direct searches for leptoquark decays into 2-body SM final states, while allowing the 3-body decays of the VALs to be prompt for a range of parameters,  as shown in Figure~\ref{fig:lambdas}.

Constraints from flavor-changing processes, such as $D^0 - \overline{D^0}$ mixing, $K^0 - \overline{K^0}$ mixing, and $K^+ \rightarrow \pi^+ \, \nu \, \overline \nu$, suggest that the Yukawa 
couplings of the new particles to first-generation quarks are suppressed. Even the couplings to second- and third-generation quarks are constrained, but only mildly (see Figure~\ref{fig:lambdas}), by 
flavor-changing processes  including $B_s-\overline B_s$ mixing, $B^0 \to K^{*0} \nu \bar \nu$, and $B^+ \to K^+ \nu \bar \nu$.

The electroweak production of vectorlike fermion pairs at the LHC, computed at NLO in $\alpha_s$ for $\sqrt{s} = 13.6$ TeV and displayed in  
Figure~\ref{fig:xsection}, applies to a VAL doublet, but also to a vectorlike lepton doublet. While the current lower mass limit is already above 1 TeV for a vectorlike lepton \cite{CMS:2022nty}, it is substantially less stringent for a VAL. If the couplings of the new particles are predominantly to the second-generation quarks, then the main LHC signals are $\tau^+\tau^- + 4j$ and $ \tau + 4j + \slash \!\! \!\! E $ for VALs, and $\tau^+\tau^- + 6j$ for leptoquarks. 
These signals with taus and jets are difficult to observe due to large backgrounds, but with advanced experimental techniques can be probed in the current Run 3 of the LHC. 

The main LHC signals  in the case of dominant couplings to third-generation quarks are 
$ \tau^\pm W^+W^- \! + 4b + \slash \!\! \!\! E $ and $\tau^+ \tau^- W^+W^- \! + 4b $ 
for VALs, and $\tau^+\tau^- + 6b + 4W$ for leptoquarks. The presence of several charged leptons from $\tau$ or $W$ decays makes these signals easier to detect than the ones with just taus and jets. Nevertheless, the large multiplicities of these signals imply that most objects in the final states are relatively soft. Thus, even in the case of third-generation couplings, the vectorlike fermions and leptoquarks make each other more elusive than in the case of standard 2-body  decays.

\bigskip\medskip\medskip

{\bf Acknowledgments:} \ We would like to thank Elias Bernreuther, Sekhar Chivukula, and Stefan H\"{o}che, for insightful comments.
Fermilab is administered by Fermi Research Alliance, LLC under Contract No. DE-AC02-07CH11359 with the U.S. Department of Energy, Office of Science, Office of High Energy Physics.

\bibliography{bibl}
\end{document}